\documentclass[12pt,draftcls, onecolumn]{IEEEtran}
\usepackage{graphics, graphicx}
\usepackage{amssymb,amsmath}
\usepackage{amsmath}
\usepackage[rm]{subfigure}    
\usepackage{threeparttable}   
\usepackage[dvips]{color}
\usepackage{url}   
\usepackage[latin1]{inputenc} 
\usepackage[T1]{fontenc}
\usepackage{algorithm}               
\usepackage{algorithmic}    
\usepackage{cite}
\usepackage{multirow}
\usepackage[dvips]{color}
\usepackage{lineno}
\usepackage[normalem]{ulem}
\usepackage{soul}  
\usepackage{indentfirst}
\usepackage{setspace}
\usepackage{cases}

\newcommand{\bm}[1]{\mbox{\boldmath{$#1$}}}

\begin{document}
\title{Total Energy Efficiency of TD- and FD-MRC Receivers for Massive MIMO Uplink}
\author{José Carlos Marinello, Cristiano Panazio, Taufik Abrão, Stefano Tomasin\\
\thanks{J. C. Marinello and T. Abrão are with the Electrical Engineering Department, State University of Londrina, PR, Brazil.  E-mail: {zecarlos.ee@gmail.com; \quad taufik@uel.br}}
\thanks{C. Panazio is with the Laboratory of Communications and Signals, Polytechnic School of the University of São Paulo, SP, Brazil. E-mail: cpanazio@lcs.poli.usp.br}
\thanks{S. Tomasin is with the Department of Information Engineering, University of Padova, Via Gradenigo 6/B, 35131, Padova, Italy. E-mail: tomasin@dei.unipd.it}
}

\maketitle 

\begin{abstract}
This paper proposes a detailed investigation on the uplink (UL) performance of massive multiple-input-multiple-output (maMIMO) systems employing maximum-ratio combining at the receiver. While most papers in maMIMO literature assume orthogonal frequency-division multiplexing (OFDM), current standards like LTE employ single-carrier (SC) waveform in the UL due to several benefits. We thus perform a systemic comparison between two fundamental schemes: the time-reversal MRC (TRMRC) operating under SC, and the  frequency-domain MRC (FDMRC) employed with OFDM. It was recently shown that TRMRC outperforms FDMRC in terms of achievable rates, since SC systems do not require the cyclic prefix (CP) of OFDM. On the other hand, the computational complexity of TRMRC algorithm is higher than that of FDMRC, even when efficient solutions are employed (e.g., fast convolution with the overlap-and-add method). Hence, the best scheme for the UL maMIMO systems still remains an open question. The main contribution of this paper is the comparison of the total energy efficiency of both TRMRC and FDMRC when used in the UL of maMIMO systems. Our results show that, for current typical system parameters, FDMRC/OFDM achieves a higher total energy efficiency than TRMRC/SC. However, if the cell radius is below 300m and/or the computational efficiency increases by 30\% regarding the current processors, the TRMRC under SC waveform becomes more attractive for the UL of maMIMO systems.
\end{abstract}


\section{Introduction}
Massive multiple-input-multiple-output (maMIMO) technology is one of the most promising candidates for the physical layer of the upcoming fifth generation of mobile communications systems (5G) \cite{Marzetta16}. It is widely known that maMIMO systems achieve very appreciable spectral and energy efficiencies \cite{Ngo13_EE_SE}, and thus they have been {the} focus of several research papers recently. As long as the number of base station (BS) antennas of a time-division duplex (TDD) multicellular system grows without bound, coherent interference is the only {limiting effect} under independent and identically distributed (i.i.d.) Rayleigh fading channels \cite{Marzetta16}. This kind of interference results from pilot contamination, which in turn results from the unavoidable reuse of reverse-link pilot sequences between users of different cells due to the limited channel coherence block. On the other hand, every type of non-coherent interference, \emph{i.e.}, thermal noise, intracell interference, and intercell interference from users with different pilots, vanishes in this scenario \cite{Marzetta16}. As a positive consequence, the simplest schemes for detection in uplink (UL) and for precoding in downlink (DL) become optimal \cite{Marzetta10}, like maximum ratio combining (MRC) and maximum ratio transmission (MRT), respectively. 

Most papers in the area of maMIMO assumes orthogonal frequency-division multiplexing (OFDM) as a way of dealing with inter-symbol interference. In the seminal maMIMO paper \cite{Marzetta10}, the author has highlighted several aspects of the proposed technology in conjunction with OFDM waveform. In \cite{Ngo13_EE_SE}, a first detailed investigation about maMIMO energy and spectral efficiencies have been conducted, assuming OFDM waveform. In \cite{Debbah16}, the authors have found the conditions in which the maximal spectral efficiency of maMIMO systems employing OFDM waveform can be achieved in different scenarios. A detailed study about power allocation schemes applied to OFDM based maMIMO systems has been conducted in \cite{Zhang15}. 

On the other hand, single-carrier (SC) waveform has been widely used in the UL of current standards, such as LTE. Several advantages of employing SC in the UL instead of OFDM can be highlighted \cite{Falconer02}, \cite{Marinello17}, such as: {\bf i}) the {transmit signal} lower peak-to-average power ratio (PAPR), which {results in cheaper and more energy efficient mobile terminals (MTs)}. Besides, this {incurs} in a {reduced} distortion level imposed by the transmit amplifier, and {consequently} in a lower out-of-band radiation, {simplifying} spectrum shaping and {increasing} spectral efficiency {due to a lower guard band required}. {\bf ii}) Lower sensitivity to phase noise and carrier frequency offset (CFO). {\bf iii}) {Reduced} complexity of the MT since the inverse fast Fourier transform (IFFT) operation is not required. {\bf iv}) {Under the} maMIMO {condition}, it is shown in \cite{Marinello17} that cyclic prefix (CP) transmission is not required employing TRMRC in conjunction with SC, while improving spectral and energy efficiencies. Due to the above mentioned benefits, in conjunction with the compatibility with the current standards, this waveform has been seen as a promising candidate for implementation in standards, such as 5G \cite{GYLi_Survey_14}.

SC receivers {normally} require complex equalization techniques in either time or frequency domain, such as decision-feedback equalizer (DFE) \cite{Benvenuto02}. {The computational complexity of} frequency domain equalization is {often lower} than {that of} its time-domain {version} {\cite{Tomasin05}}, but {usually} requiring CP transmission and one {fast Fourier transform (FFT)} module per receive antenna similarly {to} OFDM. {On the other hand}, time-domain equalization {provide a good} performance at the expense of {higher} complexity, not requiring CP and FFT operations. However, when {combined} with maMIMO, a different time-domain receiver can be {employed}, which {alleviates the equalization complexity by taking advantage of the large excess of BS antennas}, namely the time-reversal MRC (TRMRC) receiver \cite{Larsson15}. The TRMRC receiver coherently combines the received symbols arising from different paths with the appropriate channel impulse response (CIR) vector. The robustness of this scheme was investigated in \cite{Larsson15} against the phase noise effect, and a DL version of the scheme was proposed in \cite{Larsson12}, but under a scenario with no pilot contamination in both cases. 

In \cite{Mohammed16}, an improved TRMRC receiver is proposed which performs the CFO estimation/compensation together with the equalization procedure. Authors have shown that the proposed scheme performance approaches the ideal/zero CFO performance with the increasing frequency selectivity of the channel. Another DL SC solution for massive MIMO \cite{Dinis18} proposes a SC frequency domain precoding scheme. The main difference of this scheme with respect to that of \cite{Larsson12} is the frequency domain processing, which significantly reduces the computational complexity. Besides, several other papers regarding SC combine this waveform with spatial modulation (SM) \cite{Song18}, \cite{He18}, a promising multiple antenna modulation scheme which activates only a subset of the available antennas, and the indices of the activated antennas convey information. Schemes combining SM and SC have attracted much research attention recently aiming at combining the advantages of both schemes, for example lower number of required radio-frequency chains of SM, and lower PAPR of SC.

An interesting comparison of receiver/waveforms for the UL of maMIMO systems has been conducted in \cite{Marinello17} and \cite{Heath17}. In \cite{Heath17}, the UL performance of SC and OFDM receivers are compared for single cell maMIMO systems employing one-bit analog-to-digital converters (ADCs), which significantly reduce the power consumption of BSs. Both schemes have a very similar performance and one-bit ADCs work well with maMIMO systems in the investigated scenario without pilot contamination. On the other hand, under pilot contamination, the TRMRC scheme operating with SC waveform has been compared with the conventional FDMRC in conjunction with OFDM in \cite{Marinello17}. The authors have found very similar expressions for the performance of both schemes in terms of signal-to-interference-plus-noise ratio (SINR). However, {higher} achievable rates can be attained by TRMRC under SC, since this scheme does not require CP transmission, in contrast with FDMRC and OFDM. However, it was shown that the computational complexity demanded by the former is somewhat higher. Even if employing a reduced complexity algorithm with the aid of fast convolution techniques with overlap-and-add, similarly as the overlap-and-save method applied for single-input-single-output (SISO) SC systems in \cite{Tomasin05}, the complexity of the TRMRC algorithm is $\mathcal{O}(\log_2(2L)MK)$, in contrast with the ($\mathcal{O}(MK)$) of FDMRC \cite{Marinello17}. At this point, the question of what scheme consists in a better candidate for the UL of upcoming maMIMO systems remains without a complete answer, since each scheme has advantages from different perspectives.

In \cite{Debbah15} and \cite{Marzetta13}, the authors have assessed the total energy efficiency of maMIMO systems operating with OFDM waveform. In \cite{Debbah15}, authors found that the optimal energy efficiency is attained with a maMIMO setup wherein hundreds of antennas are deployed to serve a relatively large number of users if using zero forcing processing. The total energy efficiency metric computes how many bits are transmitted per Joule of energy, but taking into account every energy expenditure of the communication system. Accurate power consumption models are presented in \cite{Debbah15} for the several stages of a practical communication system, like the power consumed by the transceiver chains, the channel estimation process, the coding and decoding units, the backhaul, the linear processing algorithms, and also by the radiated signal power. 

The problem of designing the communication system aiming at maximizing the energy efficiency has attracted much research attention recently. In \cite{Yang17}, an energy-efficient resource allocation scheme is proposed for heterogeneous orthogonal frequency-division multiple access (OFDMA) networks, which allocates both the transmit power to the users as well as the OFDMA frequency resource blocks. In \cite{Fu17}, considering the coordinated multi-point transmission, the energy efficiency is maximized via block diagonalization precoding and a novel water filling power allocation algorithm with a pricing power game mechanism. Another key mechanism for increasing energy efficiency in 5G is network densification, in which the BS is brought closer to the users by deploying small cells. In \cite{Liao17}, a network architecture is proposed for ultra-dense heterogeneous networks, as well as a random access scheme able to improve the network connectivity for the increased number of accessing devices in the 5G network. Area spectral efficiency (ASE) and area energy efficiency (AEE) are important metrics for evaluating the spatial coverage of a cellular communication system. The trade-off between ASE and AEE for the UL maMIMO cellular system is investigated in \cite{Xin16}. Authors find interesting results regarding the parameters choice influence on system performance; for example, it is beneficial to employ the same  transmit power for both data and pilot transmissions. However, the analysis in \cite{Xin16} considers only uplink, while ignores the shadowing effect for analytical tractability.

In this paper, our main objective is to {\it answer which waveform between SC and OFDM is the most suitable for implementation in the maMIMO UL considering MRC receiver}. Our choice for these two waveforms is because they are the most consolidated technologies, and therefore the most prominent for implementation in future standards, contrary to, for example, filter bank multicarrier which is still in development stage. Besides, MRC does not require matrix inversion, being highly parallelizable, and asymptotically optimal in i.i.d. Rayleigh fading channels \cite{Marzetta16}. For this purpose, we evaluate the total energy efficiency of maMIMO systems employing the TRMRC technique under SC waveform in the UL, and compare the obtained results with the total energy efficiency achieved by FDMRC with OFDM. The comparison results allow us to indicate {\it which scheme (receiver/waveform) is more advantageous for implementation in the UL of maMIMO systems} since the {\it total energy efficiency metric can weight in a unified systemic analysis the better attainable rates of TRMRC against the lower complexity burden of FDMRC}. 

Our main contributions are fourfold: {\bf i}) Differently than \cite{Debbah15}, which have assessed the total energy efficiency of maMIMO systems employing OFDM waveform, we have also evaluated the overall energy efficiency of maMIMO systems using SC waveform in the UL, besides of extending the total energy efficiency concept to a more generic scenario; {\bf ii}) Differently than our previous work \cite{Marinello17}, in which the performance of maMIMO systems employing TRMRC under SC in UL is obtained only under the spectral efficiency perspective, we have evaluated the performance of such systems under the comprehensive view of total energy efficiency metric, which is also quite appealing currently aiming to reduce power consumption; {\bf iii}) Our total energy efficiency analysis and comparison allowed us to indicate the best scheme to be implemented in the maMIMO UL, since it weights both achievable rates as well as computational complexities; {\bf iv}) We have also extended the total energy efficiency comparison to scenarios with decreasing cell radius, and with increasing computational efficiency, in order to indicate the best scheme even in these different scenarios.

Besides this introductory section, the system model is described in Section \ref{sec:model}. We define the total energy efficiency of massive MIMO systems in Section \ref{sec:TotalEE}, {as well as describing our adopted realistic circuit power consumption models}. Representative numerical results comparing the uplink SC against OFDM maMIMO total energy efficiency and spectral efficiency performance are presented in Section \ref{sec:results}. Section \ref{sec:concl} concludes the paper.\\

\noindent\textit{Notations:} Boldface lower and upper case symbols represent vectors and matrices, respectively. ${\bf I}_N$ denotes the identity matrix of size $N$, while ${\bf 1}_K$ and ${\bf 0}_K$ are the unitary vector and null vector of length $K$, respectively. $\{\cdot\}^T$ and $\{\cdot\}^H$ {denotes the transpose and the Hermitian transpose operator}, respectively. {Besides,} we use $\mathcal{CN}({\bf m}, {\bf V})$ {when referring} to a circular symmetric complex Gaussian distribution with mean vector ${\bf m}$ and covariance matrix ${\bf V}$.

\section{System Model}\label{sec:model}
We consider a MIMO system composed by $C$ BSs, each equipped with $M$ transmit antennas, and communicating with $K$ single-antenna users, similarly as \cite{Marinello17}. Reciprocity holds due to the TDD assumption, and thus uplink training signals transmission provides channel state information (CSI) to the BSs. We divide {our} system model description for OFDM and SC waveforms. 

\subsection{OFDM waveform}

\begin{figure}[!htbp]
\centering
\includegraphics[width=.44\textwidth]{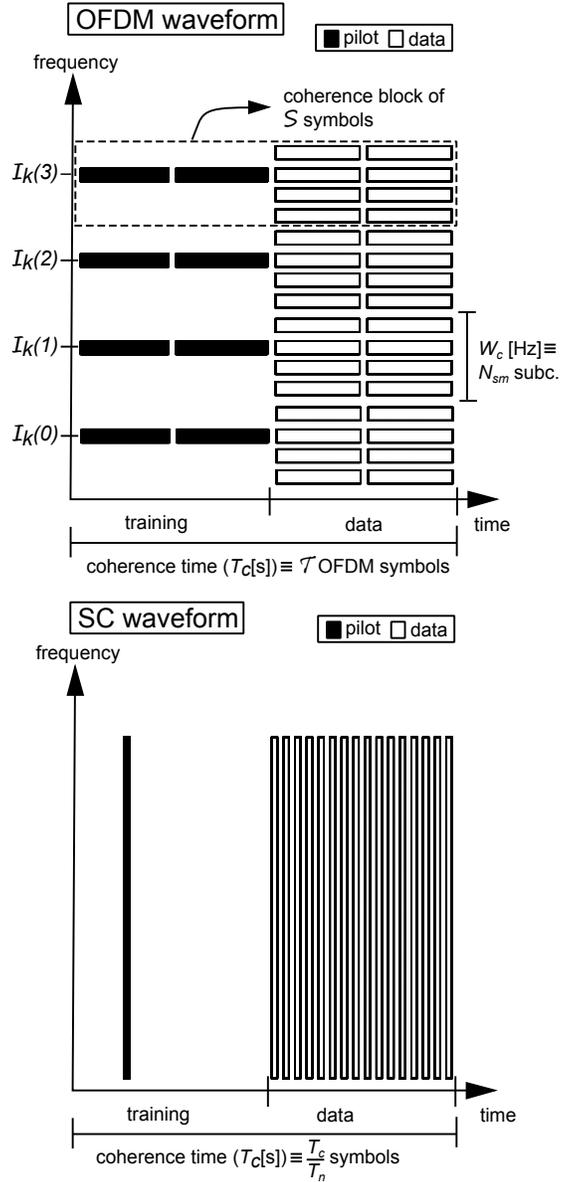}
\vspace{-2mm}
\caption{Time-frequency diagram illustrating {the division of} training and data transmission stages for the $k$th user, for OFDM and SC waveforms: $N=16$, $L = 4$, $N_{\textrm{sm}} = 4$, $\mathcal{T} = 4$, $\mathcal{S} = 16$ for OFDM; $T_c/T_n = 32$ for SC. {Besides,} $K_{\textrm{max}} = 8$ for both schemes.}
\label{fig:TDD}
\end{figure}

The {available} time-frequency resources are divided into channel coherence blocks of $\mathcal{S} = T_c W_c$ symbols \cite{Debbah16}, {where} $T_c$ is the channel coherence time in seconds, and $W_c$ is the channel coherence bandwidth in Hz\footnote{{$W_c = \frac{1}{T_d}$ holds, being $T_d$ the largest possible delay spread.}}. Each coherence time interval {comprises} $\mathcal{T}$ OFDM symbols ($T_c = \mathcal{T}\,T_s$, {with $T_s$ representing} the OFDM symbol period), distributed as uplink pilot transmissions, downlink and uplink data transmissions \cite{Marzetta10}. {The number of available orthogonal pilot} sequences is equal to its length, $\tau$. In the same way as \cite{Marzetta10} and \cite{Marinello17}, CSI is estimated in the frequency domain. With the system {employing} $N$ subcarriers, and the CIR {being represented by} $L$ i.i.d. taps, the frequency smoothness interval is composed of $N_{\textrm{sm}} = N/L$ subcarriers. {Therefore, it is not required to each user to send} pilots in all the $N$ subcarriers, but only in $N/N_{\textrm{sm}} = L$ equally spaced subcarriers, since {a interpolation procedure can provide the frequency domain CSI (FDCSI) for the intermediate ones}. {In the training stage,} up to $\tau$ users can send pilots in the same subcarriers, {while} a maximum total number of $K_{\textrm{max}} = \tau N_{\textrm{sm}}$ users can be {covered} by each cell. {The set of subcarriers in which the $k$th user sends pilots is defined} as $\bm{\mathcal{I}}_{k} = \{\mathcal{I}_{k}(0), \mathcal{I}_{k}(1), \ldots, \mathcal{I}_{k}(L-1)\}$, {while} $\mathcal{I}_{k}(0) \in [0, N_{\textrm{sm}}-1]$ and $\mathcal{I}_{k}(i) - \mathcal{I}_{k}(i-1) = N_{\textrm{sm}}, \forall i = 0, 1, \ldots, L-1$. {Fig.} \ref{fig:TDD} {outlines} how training and data transmission stages are divided in time and frequency domains for a given user, {being} $T_n$ the Nyquist sampling {interval}.

The $M \times 1$ vector ${\bf g}_{i k j n}$ denotes the FD actual channel between the {$i$}th BS and the $k$th user at the $j$th cell in the $n$th subcarrier. {It holds that ${\bf g}_{i k j n} = \sqrt{\beta_{i k j}} {\bf \underline{g}}_{i k j n}$}, in which $\beta_{i k j}$ is the long-term fading power coefficient, {comprising} path loss and log-normal shadowing, and ${\bf \underline{g}}_{i k j n}$ is the short-term fading channel vector, {following} ${\bf \underline{g}}_{i k j n} \sim \mathcal{CN}({\bf 0}_M, {\bf I}_M)$. The orthogonal pilot sequences' set is ${\bm \Psi} = [{\bm \psi}_1  \,\, {\bm \psi}_2 \, \ldots \, {\bm \psi}_\tau] \in \mathbb{C}^{\tau\times \tau}$, and ${\bm \Psi}^H {\bm \Psi} = {\bf I}_{\tau}$ holds. With the sequence ${\bm \psi}_k = [\psi_{k 1} \,\, \psi_{k 2} \ldots \psi_{k \tau}]^{T}$ assigned {to} the $k$th user, the received signal at the $i$th BS for the $n$th subcarrier ($n \in \bm{\mathcal{I}}_{k}$) during the training stage is
\begin{equation}\label{eq:rx_pilots}
{\bf Y}^{\rm p}_{i n} = \sqrt{\rho_{\rm p}} \sum_{j = 1}^{C} \sum_{k = 1}^{\tau} {\bf g}_{ikjn} {\bm \psi}^H_k + {\bf N}^{\rm p}_{i n},
\end{equation}
{with} $\rho_{\rm p}$ {representing} the UL pilot transmit power, {and} ${\bf N}^{\rm p}_{i n} \in \mathbb{C}^{M \times \tau}$ the {additive white Gaussian noise (AWGN)} matrix with i.i.d. elements following a complex normal distribution with zero mean and variance $\sigma^2_n$. By correlating the received signal ${\bf Y}^{\rm p}_{i n}$ with ${\bm \psi}_k$, {the $i$th BS then estimates the $k$th user $n$th subcarrier FDCSI as}
\begin{equation}\label{eq:rx_correlator}
\widehat{{\bf g}}_{ikn} = \frac{1}{\sqrt{\rho_{\rm p}}} {\bf Y}^{\rm p}_{i n} {\bm \psi}_k = \sum_{j = 1}^{C} {\bf g}_{ikjn}  + {\bf v}_{ikn},
\end{equation}
where ${\bf v}_{ikn} = \frac{1}{\sqrt{\rho_{\rm p}}} {\bf N}^{\rm p}_{i n} {\bm \psi}_k \sim \mathcal{CN}({\bf 0}_M, \frac{\sigma^2_n}{\rho_{\rm p}} {\bf I}_M)$ is an equivalent noise vector. {The previous expression clearly demonstrates the pilot contamination effect \cite{Marzetta10}}. The other $N-L$ subcarriers FDCSI {estimates} are obtained by interpolation. The BS is {now} able to perform linear detection in the UL deploying {the} FDMRC scheme. During UL data transmission, the $i$th BS receives in the $n$th subcarrier the signal
\begin{equation}\label{eq:rx_data_u}
{\bf y}^{\rm u}_{i n} = \sqrt{\rho_{\rm u}} \sum_{j = 1}^{C} \sum_{k = 1}^{K} {\bf g}_{ikjn} x^{\rm u}_{k j n} + {\bf n}^{\rm u}_{i n},
\end{equation}
{where} $\rho_{\rm u}$ {represents} the UL data transmit power, $x^{\rm u}_{k j n}$ is the data symbol from the $k$th user of the $j$th cell in the $n$th subcarrier, and ${\bf n}^{\rm u}_{i n} \sim \mathcal{CN}({\bf 0}_M, \sigma^2_n {\bf I}_M)$ is the $M\times 1$ AWGN sample vector. This BS then {deploys} FDMRC {to obtain the estimated transmitted symbol} as 
\begin{equation}\label{eq:est_data_fdmrc}
\widehat{x}^{\rm u}_{k i n} = \frac{1}{M} \widehat{{\bf g}}_{ikn}^H {\bf y}^{\rm u}_{i n}.
\end{equation}

\subsection{SC waveform}\label{sec:SCwavef}

The $t$th CIR vector tap between the $i$th BS and the $k$th user of the $j$th cell and its FD actual channel vector are related by
\begin{equation}\label{eq:fdcsi_cir}
{\bf h}_{ikjt} = \frac{1}{L} \sum_{l=0}^{L-1} {\bf g}_{ikj\mathcal{I}_{k}(l)} e^{\texttt{j} 2 \pi l t/L}.
\end{equation}

We assume in this work a normalized uniform power delay profile (PDP) for all users, such that ${\bf h}_{ikjt} \sim \mathcal{CN}({\bf 0}_M, \frac{1}{L} {\bf I}_M)$ for $t = 0, 1, \ldots L-1$, {since this is the most generic scenario}. {When} employing single-carrier waveform, {a training scheme for estimating the CIR of the users served by the cell is discussed in \cite{Marinello17} and \cite{Larsson15}, which is alternative to conventional frequency domain training}. In this alternative scheme, the users transmit uplink training signals sequentially in time, in such a way that at most one user per cell is transmitting at any given time. The $k$th user sends an impulse of amplitude $\sqrt{L \rho_{\rm p}}$ (in order to spend the same average power as in FD training \eqref{eq:rx_pilots}) at the $(k-1)L$th channel use and is idle for the remaining portion of the training phase. Besides, the next user waits $L-1$ channel uses to send its training signal. It is worth noting that the same maximal number of users $K_{\textrm{max}}$ is supported in this scheme. This can be proved by noting that the total interval spent with training signals is $K_{\textrm{max}} L \, T_{n} = \tau N_{\textrm{sm}} L \, T_{n}= \tau N \, T_{n} \approx \tau T_s$ is nearly the same as in FD training, as illustrated in {Fig.} \ref{fig:TDD}. {This scheme provides the channel estimation in time domain, avoiding the transmission of FD pilots and the use of FFT at the receiver, not otherwise required in SC transmission \cite{Larsson15}.}

This procedure provides to the $i$th BS the following CIR estimates for its $k$-th user
\begin{equation}\label{eq:rx_correlator_td}
\widehat{{\bf h}}_{ikt} = \sum_{j = 1}^{C} {\bf h}_{ikjt}  + {\bf v}'_{ikt},
\end{equation}
in which ${\bf v}'_{ikt} \sim \mathcal{CN}({\bf 0}_M, \frac{\sigma^2_n}{L \rho_{\rm p}} {\bf I}_M)$ is an equivalent AWGN vector. Note that {pilot contamination also takes place in \eqref{eq:rx_correlator_td}, since} the CIR estimates are also corrupted by the {signals of users} in adjacent cells {transmitting} at the same {time}.

Considering SC waveform, the time-domain received signal at the $i$th BS in time $t$ during UL data transmission is
\begin{equation}\label{eq:rx_data_u_td}
{\bf r}^{\rm u}_{i}[t] = \sqrt{\rho_{\rm u}} \sum_{j = 1}^{C} \sum_{k = 1}^{K} \sum_{l = 0}^{L-1} {\bf h}_{ikjl} s^{\rm u}_{k j}[t - l] + {\bf n'}^{\rm u}_{i}[t],
\end{equation}
in which $s^{\rm u}_{k j}[t - l]$ is the time-domain UL data transmitted by the $k$th user of the $j$th cell at time $t - l$, and ${\bf n'}^{\rm u}_{i}[t] \sim \mathcal{CN}({\bf 0}_M, \sigma^2_n {\bf I}_M)$ is an AWGN vector. Note that the time aligned received signals in \eqref{eq:rx_data_u_td} are resultant from a perfect timing advance assumed at the users, as in \cite{Marinello17}, \cite{Heath17}. The BS can then estimate the {information} symbols deploying TRMRC as
\begin{equation}\label{eq:est_data_trmrc}
\widehat{s \,}^{\rm u}_{k i}[t] = \frac{1}{M} \sum_{l=0}^{L-1} \widehat{{\bf h}}_{ikl}^H {\bf r}^{\rm u}_{i}[t+l].
\end{equation}

For causality purpose, {\eqref{eq:est_data_trmrc} demonstrates that} $L-1$ shift registers are necessary per BS receive antenna {when employing} TRMRC, while this number for FDMRC is $N-1$, since the entire frame has to be received to perform FFT operation. This implies not only hardware simplification, but also latency reduction. 
Besides, TRMRC {does not require} CP {as} seen from \eqref{eq:est_data_trmrc}, since given a sufficiently large $M$, the BS extracts the desired information symbol vector from ${\bf r}^{\rm u}_{i}[t+l]$, even if the asynchronous interference belongs to the previous frame of symbols \cite{Marinello17}. This is in fact one of the main advantages of TRMRC in comparison with FDMRC, which comes, however, at the cost of increased computational complexity of the time domain processing, as discussed in Section \ref{sec:complexity}.

\subsection{System Performance}\label{sec:SystPerf}

The analytical SINR performance of the $k'$th user at the $i$th cell achieved by FDMRC and TRMRC schemes have been found in \cite{Marinello17} as 
\begin{eqnarray}\label{eq:sinr_fdmrc}
\varsigma^{\textsc{f}}_{k' i} = \frac{M \beta^2_{ik'i}}{M \sum_{\substack{c = 1 \\ c \neq i}}^{C} \beta^2_{ik'c} + \alpha^2_{i k'} \left[\sum_{j = 1}^{C} \sum_{k=1}^{K} \beta_{ikj} + \frac{\sigma_n^2}{\gamma \rho_{\rm u}}\right]}
\end{eqnarray}
{and}
\begin{eqnarray}\label{eq:sinr_trmrc}
\varsigma^{\textsc{t}}_{k' i} = \frac{M \beta^2_{ik'i}}{M \sum_{\substack{c = 1 \\ c \neq i}}^{C} \beta^2_{ik'c} + \alpha^2_{i k'} \left[\sum_{j = 1}^{C} \sum_{k=1}^{K} \beta_{ikj} + \frac{\sigma_n^2}{\rho_{\rm u}}\right]},
\end{eqnarray}
in which $\alpha^2_{i k'} = \sum_{c=1}^{C} \beta_{ik'c} + \frac{\sigma_n^2}{\rho_{\rm p}}$. The parameter $\gamma = \frac{N}{N+N_{CP}}$ accounts for the transmit power loss due to CP, since we have assumed for a fair comparison that both schemes spend the same energy with data transmission, and $N_{CP} = L-1$ is the CP length. Note that this result is valid for all subcarriers, since they are identically distributed in this model.

{\it Remark 1:} One can note from \eqref{eq:sinr_fdmrc} and \eqref{eq:sinr_trmrc} that the same SINR performance can be achieved by both TRMRC/SC and FDMRC/OFDM schemes, except by the decrease in the transmit power of the latter due to the CP transmission. Besides, analysing both expressions, we note that the desired signal power is mainly dependent on the large-scale fading coefficient of the user, and scales with $M$. The first interference term is due to the users in adjacent cells sharing the same pilot sequence, and is also called as coherent interference, since the receiver processing is not able to mitigate this term which also scales with $M$. The second term in the denominator involves interference from users with different pilots and noise, and is so-called non-coherent interference, since the receiver processing is able to mitigate this term, for example with more BS antennas.

Besides, due to the convexity of $\log_2(1+\frac{1}{x})$ and using Jensen's inequality as in \cite{Ngo13_EE_SE}, we obtain {for FDMRC} the following lower bound on the achievable rate of the $k'$th user at the $i$th cell
\begin{eqnarray}\label{eq:se_fdmrc}
\mathcal{R}^{\textsc{f}}_{k' i} \geq \widetilde{\mathcal{R}}^{\textsc{f}}_{k' i} = \xi^{\rm u} \gamma \left(1 - \frac{K}{\mathcal{S}} \right) \log_2\left( 1 + \left( \frac{1}{\varsigma^{\textsc{f}}_{k' i}} \right)^{-1} \right),
\end{eqnarray}
in which the term $\left(1 - \frac{K}{\mathcal{S}} \right)$ accounts for the pilot transmission overhead, and $\xi^{\rm u}$ is the fraction of the data transmit interval spent with uplink.

For TRMRC we have similarly:
\begin{eqnarray}\label{eq:se_trmrc}
\mathcal{R}^{\textsc{t}}_{k' i} \geq \widetilde{\mathcal{R}}^{\textsc{t}}_{k' i} = \xi^{\rm u} \left(1 - \frac{K}{\mathcal{S}} \right) \log_2\left( 1 + \left( \frac{1}{\varsigma^{\textsc{t}}_{k' i}} \right)^{-1} \right).
\end{eqnarray}

Furthermore, the asymptotic UL performance of both systems converge with infinite number of BS antennas to the same bound of \cite{Marzetta10}:

\begin{equation}\label{eq:asymptSINR}
\lim_{M \to \infty} \varsigma^{\textsc{f}}_{k' i} = \lim_{M \to \infty} \varsigma^{\textsc{t}}_{k' i} = \frac{\beta^2_{ik'i}}{\sum_{\substack{c = 1 \\ c \neq i}}^{C} \beta^2_{ik'c}}.
\end{equation}

\subsection{Complexity Analysis}\label{sec:complexity}

The computational complexity of both investigated schemes is evaluated in \cite{Marinello17}, in terms of \emph{floating-point operations (flops)} \cite{Golub96}. The main assumptions that have been made are: {\it i}) A real sum operation has the same complexity than a real multiplication, which are defined as 1 \emph{flop}\footnote{It is important to note that the exact factor between the complexities of sum and multiplication operations does not play a significant role in our results, since in a complexity analysis, the most significant operations dominate the final complexity expressions; our analysis is comparative, and thus changing this relative factor would impact on the complexities of both schemes in a similar manner; in the total energy efficiency analysis the computational processing energy accounts for only part of the total energy expenditures. Although some works \cite{Horowitz17} \cite{Horowitz14} have shown that the energy of the multiplication operation between two double precision floating point numbers is about four times the energy of the addition operation, the memory access energy may be dominant and similar for both operations.}; {\it ii}) A complex sum has the complexity of 2 \emph{flops}, while a complex multiplication has the complexity of 6 \emph{flops}; {\it iii}) An inner product between 2 complex vectors of size $M$ has the complexity of $M$ complex multiplications and $M-1$ complex additions, resulting in a total complexity of $8M-2$ \emph{flops}; {\it iv}) A FFT or IFFT operation of a complex sequence of $N$ samples has the complexity (supposing $N$ power of two and adopting the Cooley Tukey algorithm \cite{Golub96}) of $(N/2) \log_2(N)$ complex multiplications and $N \log_2(N)$ complex additions, and an overall complexity of $5N \log_2(N)$ \emph{flops}. {\it v}) Finally, the complexity related with the CSI estimation has been neglected for all the investigated schemes, since these estimates remain valid for several frame periods, and account in the same way for all the schemes.

For the FDMRC under OFDM, the complexity to detect the $N$ symbols of the OFDM frame is mainly due to: {\it i}) the $N$-points IFFT operations of each user transmission ($K [5 N \log_2(N)]$ \emph{flops}), {\it ii}) the $N$-points FFT operations of each BS receive antenna ($M [5 N \log_2(N)]$ \emph{flops}), and due to {\it iii}) the $N$ FDMRC data symbols' estimation, as in \eqref{eq:est_data_fdmrc}, for each user ($N K [8 M -2]$ \emph{flops}). On the other hand, the complexity of the TRMRC with overlap-and-add method is mainly given by: {\it i}) $\mu = (N+L)/L$ FFT's of size $2L$ for the received signals of each receive antenna, resulting in a complexity of $M \mu [10 L \log_2(2L)]$ \emph{flops}; {\it ii}) For each user and each receive antenna: $\mu$ element-wise multiplication of size $2L$ vectors, followed by an IFFT operation of size $2L$, and the addition of the overlaping segments (of size $L$) ($K M \mu [10 L \log_2(2L) + 12 L + 2 L]$ \emph{flops}); {\it iii}) For each user, the results of the $M$ fast convolutions are added (only the $N$ elements of interest), with a complexity of $K (2 N (M-1))$ \emph{flops}. The overall complexity of the TRMRC receiver employing the fast convolution with overlap-and-add method is also given in Table \ref{tab:complexity}.

\footnotesize
\begin{table}[!htbp]
\caption{Number of Operations (flops) for each scheme.}
\vspace{-2mm}
\centering
\small
\begin{tabular}{|c|c|}
\hline
\bf Scheme & \bf Number of Operations\\
\hline
\hline
FDMRC & $K [5 N \log_2(N)] + M [5 N \log_2(N)]$\\
 & $ + N K [8 M - 2]$\\
\hline
TRMRC & $K [ M (N+L) (10 \log_2 (2L) + 14) + 2 N (M-1) ]$\\
 & $+ M (N+L) 10 \log_2 (2L) $\\
\hline
\end{tabular}
\label{tab:complexity}
\end{table}
\normalsize

\section{Total Energy Efficiency}\label{sec:TotalEE}
We define in this subsection the {\it total energy efficiency} of a maMIMO system employing either FDMRC or TRMRC in the UL. Our main objective with this analysis is to identify the best scheme to be implemented in the upcoming maMIMO systems based on a comprehensive analysis. The choice of the total energy efficiency metric is due to its ability of incorporating different aspects of each investigated topology, making the analysis more comprehensive and fair. The higher achievable rates of TRMRC, the reduced complexity of FDMRC, and also the consequent better power efficiency of the power amplifier in a SC scheme due to the lower PAPR levels are examples of different features that are incorporated on this analysis.

The total energy efficiency metric is defined as
\begin{equation}\label{eq:TotalEE}
{\rm EE} = \frac{\sum_{k=1}^{K}\left( \mathcal{R}^{\textsc{ul}\star}_{k i} + \mathcal{R}^{\textsc{dl}}_{k i} \right)}{P_{\textsc{tx}}^{\textsc{ul}\star}+P_{\textsc{tx}}^{\textsc{dl}}+P_{\textsc{tx}}^{{\text{tr}}\star}+P^\star_{\textsc{cp}}}.
\end{equation}

The superscript $\{\cdot\}^\star$ is used in order to indicate that a given parameter depends on which waveform/detector has been employed in UL. In \eqref{eq:TotalEE}, $\mathcal{R}^{\textsc{ul}\star}_{k i}$ constitutes the UL rate of the $k$th user of the $i$th cell, and can be found by \eqref{eq:se_fdmrc} or \eqref{eq:se_trmrc}, \emph{i.e.}, $\mathcal{R}^{\textsc{ul}\star}_{k i} = \mathcal{R}^{\textsc{f}}_{k i}$ or $\mathcal{R}^{\textsc{ul}\star}_{k i} = \mathcal{R}^{\textsc{t}}_{k i}$. Besides, $\mathcal{R}^{\textsc{dl}}_{k i}$ is the DL rate of this user. We assume that a MRT precoding has been employed in downlink under OFDM waveform, since the practical appeal for SC usage in DL is not so strong as it is in UL, and thus $\mathcal{R}^{\textsc{dl}}_{k i}$ can be found {similarly to \eqref{eq:se_fdmrc}} as:
\begin{eqnarray}\label{eq:se_DL}
\mathcal{R}^{\textsc{dl}}_{k' i} \geq \widetilde{\mathcal{R}}^{\textsc{dl}}_{k' i} = \xi^{\rm d} \gamma \left(1 - \frac{K}{\mathcal{S}} \right) \log_2\left( 1 + \left( \frac{1}{\varsigma^{\textsc{dl}}_{k' i}} \right)^{-1} \right),
\end{eqnarray}
in which  $\xi^{\rm d}$ is the fraction of the data transmit interval spent with downlink, and
\begin{eqnarray}\label{eq:SINR_DL}
\varsigma^{\textsc{dl}}_{k' i} = \frac{\beta^2_{i k' i}/\alpha^2_{i k'}}{\sum^{C}_{\substack{l = 1 \\ l \neq i}} \beta^2_{l k' i}/\alpha^2_{l k'} + \frac{1}{M} \left(K \sum_{l=1}^{C} \beta_{l k' i} + \frac{\sigma^2_n}{\rho^{\rm d}} \right)}
\end{eqnarray}
is the DL SINR of the $k'$th user of the $i$th cell, while $\rho^{\rm d}$ is the uniform DL transmit power.

Furthermore, $P_{\textsc{tx}}^{\textsc{ul}\star}$, $P_{\textsc{tx}}^{\textsc{dl}}$, and $P_{\textsc{tx}}^{{\text{tr}}\star}$ in \eqref{eq:TotalEE} represent the power consumed by the power amplifiers (PA) in UL, DL, and UL pilot transmission, respectively. In our scenario of uniform power allocation {we have}
\begin{equation}
P_{\textsc{tx}}^{\textsc{ul}\star} = \frac{K \xi^{\rm u} \left( 1 - \frac{K}{\mathcal{S}} \right) \rho^{\rm u}}{\eta^{{\rm u}\star}},
\end{equation}
\begin{equation}
P_{\textsc{tx}}^{\textsc{dl}} = \frac{K \xi^{\rm d} \left( 1 - \frac{K}{\mathcal{S}} \right) \rho^{\rm d}}{\eta^{\rm d}},
\end{equation}
\begin{equation}
P_{\textsc{tx}}^{{\text{tr}}\star} = \frac{K \left( \frac{K}{\mathcal{S}} \right) \rho^{\rm p}}{\eta^{{\rm u}\star}},
\end{equation}
in which $\eta^{{\rm u}\star}$ and $\eta^{\rm d}$ represent the PA efficiency at the MT's and at the BS's, respectively. The parameter $\eta^{{\rm u}\star}$ is specific for each topology chosen as waveform/detector, since a PA with better eficiency can be employed in the MT's if SC is adopted instead of OFDM.

Finally, $P^\star_{\textsc{cp}}$ accounts for the circuit power consumption, which is discussed in detail in the following.

\subsection{Realistic Circuit Power Consumption Model}\label{sec:PowerCons}
We discuss in this {subsection} realistic models for the circuit power consumption of the different stages in a practical MIMO communication system. The adopted approach is similar to that of \cite{Debbah15}. Adopting a realistic circuit power consumption model, the parameter $P^\star_{\textsc{cp}}$ is not fixed, but dependent on several factors, like number of antennas, number of users, waveform employed, adopted precoding and detector schemes, data throughput, among others. Following \cite{Debbah15}, a refined circuit power consumption model for multi-user MIMO systems can be stated as
\begin{equation}\label{eq:cp_model}
P^\star_{\textsc{cp}} = P_{\textsc{fix}} + P_{\textsc{tc}} + P_{\textsc{ce}} + P^\star_{\textsc{c/d}} + P^\star_{\textsc{bh}} + P^\star_{\textsc{lp}},
\end{equation}
in which $P_{\textsc{fix}}$ is a constant quantity accounting for the fixed power consumption required for site-cooling, control signaling, and load-independent power of backhaul infrastructure and baseband processors \cite{Debbah15}. The other terms in \eqref{eq:cp_model} are defined in the following subsections.

\subsubsection{Transceiver Chains}

In \eqref{eq:cp_model}, $P_{\textsc{tc}}$ accounts for the power consumption of the transceiver chains. As described in \cite{Debbah15}, \cite{Bahai04}, one can quantify the power consumption of a set of transmitters and receivers as
\begin{equation}\label{eq:tc_model}
P_{\textsc{tc}} = P_{\textsc{syn}} + M P_{\textsc{bs}} + K P_{\textsc{mt}},
\end{equation}
in which $P_{\textsc{syn}}$ is the power consumed by the local oscillator, $P_{\textsc{bs}}$ is the power necessary to the circuit components of each BS antenna operate, like converters, mixers, and filters, while $P_{\textsc{mt}}$ accounts for the power required by the circuit components of each single-antenna MT operate, including amplifiers, mixer, oscillator and filters.

\subsubsection{Channel Estimation}

$P_{\textsc{ce}}$ accounts for the power consumption of channel estimation in \eqref{eq:cp_model}. We assume only UL training by UL pilot transmissions. Thus, all processing is carried out at the BS, which has a {\it computational efficiency} of $\mathcal{L}_{\textsc{bs}}\, \left[\frac{\rm flops}{\rm Watt}\right]$. The pilot-based CSI estimation is performed once per coherence block, and there are $B/\mathcal{S}$ coherence blocks per second, being $B = 1/T_n$ the total bandwidth. The CSI estimation procedure as described in \eqref{eq:rx_correlator} consists of simply multiplying the $M \times \tau$ matrix ${\bf Y}^{\rm p}_{i n}$ with the $\tau \times 1$ pilot sequence ${\bm \psi}_k$ of each user, demanding thus $M K (8 \tau - 2)$ \emph{flops}. Therefore, the power consumption spent with the channel estimation procedure is \cite{Debbah15}
\begin{equation}\label{eq:ce_model}
P_{\textsc{ce}} = \frac{B}{\mathcal{S}} \frac{M K (8 \tau - 2)}{\mathcal{L}_{\textsc{bs}}}.
\end{equation}

\subsubsection{Coding and Decoding}
In \eqref{eq:cp_model}, $P^\star_{\textsc{c/d}}$ accounts for the power consumption of the channel coding and decoding units. The BS performs channel coding and modulation to $K$ sequences of information symbols in the DL, while each MT performs some suboptimal fixed-complexity algorithm for decoding its own sequence \cite{Debbah15}. The opposite is performed in the UL. The consumed power on these procedures will thus {depend} on the actual rates, {being} expressed as
\begin{equation}\label{eq:cd_model}
P^\star_{\textsc{c/d}} = \sum_{k=1}^{K} \left( \mathcal{R}^{\textsc{ul}\star}_{k i} + \mathcal{R}^{\textsc{dl}}_{k i} \right) (\mathcal{P}_{\textsc{cod}}+\mathcal{P}_{\textsc{dec}}).
\end{equation}
where $\mathcal{P}_{\textsc{cod}}$ and $\mathcal{P}_{\textsc{dec}}$ are the coding and decoding {\it power densities}, respectively, in $\left[\frac{\rm Watt}{\rm bit/s}\right]$. In the same way as \cite{Debbah15}, it was assumed for simplicity that the power densities $\mathcal{P}_{\textsc{cod}}$ and $\mathcal{P}_{\textsc{dec}}$ are the same in UL and DL.

\subsubsection{Backhaul}

$P^\star_{\textsc{bh}}$ accounts for the power consumption of the load-dependent backhaul in \eqref{eq:cp_model}, which is necessary to transfer UL/DL data between the BS and the core network. The backhaul consumption power is usually modeled with two components \cite{Debbah15}, one load-independent and one load-dependent. For convenience, the first one can be included in $P_{\textsc{fix}}$, while the second one is proportional to the average sum rate. Therefore, the load-dependent backhaul consumption power can be computed as
\begin{equation}\label{eq:bh_model}
P^\star_{\textsc{bh}} = \sum_{k=1}^{K} \left( \mathcal{R}^{\textsc{ul}\star}_{k i} + \mathcal{R}^{\textsc{dl}}_{k i} \right) \mathcal{P}_{\textsc{bt}},
\end{equation}
in which $\mathcal{P}_{\textsc{bt}}$ is the {\it backhaul traffic power density} in  $\left[\frac{\rm Watt}{\rm bit/s}\right]$.

\subsubsection{Linear Processing}

In \eqref{eq:cp_model}, $P^\star_{\textsc{lp}}$ accounts for the power consumption of the linear processing at the BS. This power component can be divided for UL and DL. The power spent with the DL linear processing, which is a common factor for both investigated schemes, is represented as the sum of two terms: the first one is that required by making one matrix-vector multiplication per DL data symbol, and the second one is that required to obtain the precoding matrix. Each multiplication between a $M\times K$ precoding matrix and a $K\times 1$ symbol vector demands the complexity of  $M (8K-2)$ \emph{flops}. Besides, in order to obtain the MRT precoders, it is just necessary to normalize the CSI estimate vectors aiming to obtain the precoding vector of each user, demanding thus $K (14 M - 2)$ \emph{flops}. Therefore, the power consumption of the DL linear processing at the BS can be quantified as
\begin{equation}\label{eq:lpdl_model}
P^{\textsc{dl}}_{\textsc{lp}} = B \left( 1 - \frac{\tau}{\mathcal{S}} \right) \xi^{\rm d} \frac{M (8K-2)}{\mathcal{L}_{\textsc{bs}}} + \frac{B}{\mathcal{S}} \frac{K (14 M - 2)}{\mathcal{L}_{\textsc{bs}}}.
\end{equation}

The computational complexity of UL linear processing when employing FDMRC scheme under OFDM waveform is presented in Table \ref{tab:complexity}. As the first term refers to a processing evaluated at the MT's, while the others refer to BS processing, this power can be quantified as
\begin{eqnarray}\label{eq:lpulf_model}
P^{\textsc{ul-f}}_{\textsc{lp}} =& B \left( 1 - \frac{\tau}{\mathcal{S}} \right) \xi^{\rm u} \left[ K \frac{5N \log_2(N)}{\mathcal{L}_{\textsc{mt}}} + M \frac{5N \log_2(N)}{\mathcal{L}_{\textsc{bs}}} +\right. \notag\\
& \left. + N \frac{K (8 M - 2)}{\mathcal{L}_{\textsc{bs}}} \right],
\end{eqnarray}
in which $\mathcal{L}_{\textsc{mt}}$  is the computational efficiency of MT's in $\left[\frac{\rm flops}{\rm Watt}\right]$. Note that, different than \cite{Debbah15}, we included the power required to the MT's perform OFDM modulation, and the power required to the BS's perform OFDM demodulation. The computational complexity of UL linear processing when employing TRMRC scheme under SC waveform is also presented in Table \ref{tab:complexity}. As this processing is entirely evaluated at the BS, the power can be quantified as
\begin{eqnarray}\label{eq:lpult_model}
P^{\textsc{ul-t}}_{\textsc{lp}} =& \hspace{-2mm} B \left( 1 - \frac{\tau}{\mathcal{S}} \right) \frac{\xi^{\rm u}}{\mathcal{L}_{\textsc{bs}}} \left\{ K [ M (N+L) (10 \log_2 (2L) + 14)    \right. \notag \\
& \left. + 2 N (M-1)] + M (N+L) 10 \log_2 (2L) \right\} 
\end{eqnarray}

Finally, the total power required by linear processing is given by
\begin{equation}\label{eq:lp_model}
P^\star_{\textsc{lp}} = P^{\textsc{dl}}_{\textsc{lp}} + P^{\textsc{ul}\star}_{\textsc{lp}},
\end{equation}
in which $P^{\textsc{ul}\star}_{\textsc{lp}} = P^{\textsc{ul-f}}_{\textsc{lp}}$ or $P^{\textsc{ul}\star}_{\textsc{lp}} = P^{\textsc{ul-t}}_{\textsc{lp}}$.

\footnotesize
\begin{table*}[!htbp]
\caption{Simulation Parameters.}
\vspace{-2mm}
\centering
\small
{\renewcommand{\arraystretch}{1.4}%
\begin{tabular}{|l|l|}
\hline
\bf Parameter & \bf Value\\
\hline
\hline
Cell radius: $r_{\rm cell}$ & 500 m\\
 \hline
Minimum distance:  $d_{\min}$ & 50 m\\
 \hline
Transmission bandwidth: $B$ & 20 MHz\\
 \hline
Channel coherence bandwidth: $B_C$ & 100 kHz\\
 \hline
Channel coherence time: $T_C$ & 2 ms\\
 \hline
Total noise power: $\sigma_n^2$ & $-96$ dBm\\
 \hline
UL transmit power: $\rho_{\rm u}$ & 200 mW\\
 \hline
UL pilot transmit power: $\rho_{\rm p}$ & 200 mW\\
 \hline
DL radiated transmit power: $P_{\textsc{tx}}^{\textsc{dl}} \cdot \eta^{\rm d}$ & 2 W\\
 \hline
Coherence block: $\mathcal{S}$ & 200 symbols\\
 \hline
Computational efficiency at BSs: $\mathcal{L}_{\textsc{bs}}$ & $12.8 \, \left[\frac{\rm Gflops}{\rm W}\right]$\\
 \hline
Computational efficiency at MTs: $\mathcal{L}_{\textsc{mt}}$ & $5.0 \, \left[\frac{\rm Gflops}{\rm W}\right]$\\
 \hline
Fraction of DL transmission: $\xi^{\rm d}$ & 0.60\\
 \hline
Fraction of UL transmission: $\xi^{\rm u}$ & 0.40\\
 \hline
PA efficiency at the BSs: $\eta^{\rm d}$ & 0.39\\
 \hline
PA efficiency at the MTs under OFDM: $\eta^{{\rm u} \textsc{f}}$ & 0.30\\
 \hline
PA efficiency at the MTs under SC: $\eta^{{\rm u} \textsc{t}}$ & 0.50\\
 \hline
Fixed power consumption: $P_{\textsc{fix}}$ & 18 W\\
 \hline
Power consumed by local oscillators at BSs: $P_{\textsc{syn}}$ & 2 W\\
 \hline
Power required to run the circuit components at BSs: $P_{\textsc{bs}}$ & 1 W\\
 \hline
Power required to run the circuit components at MTs: $P_{\textsc{mt}}$ & 0.10 W\\
 \hline
Power density required for coding of data signals: $\mathcal{P}_{\textsc{cod}}$ & $0.10 \, \left[\frac{\rm W}{\rm Gbits}\right]$\\
 \hline
Power density required for decoding of data signals: $\mathcal{P}_{\textsc{dec}}$ & $0.80 \, \left[\frac{\rm W}{\rm Gbits}\right]$\\
 \hline
Power density required for backhaul traffic: $\mathcal{P}_{\textsc{bt}}$ & $0.25 \, \left[\frac{\rm W}{\rm Gbits}\right]$\\
 \hline
\end{tabular}}
\label{tab:parameters}
\end{table*}
\normalsize

\subsection{Other Issues Affecting Total Energy Efficiency}\label{sec:OtherIssues}
We discuss in this subsection the influence of other factors which can impact on the total energy efficiency of the investigated systems. The PAPR is an impairment which affects much more a multicarrier system like OFDM than SC systems. Increasing the number of subcarriers, $N$, has the positive effect of diminishing the inefficiency of the CP transmission, since $\gamma = \frac{N}{N + N_{CP}}$. However, it also increases the PAPR, which either requires a power amplifier with a larger linear operation range and thus lower energy efficiency, or increases the distortion levels if the PAPR is controlled by clipping the signal peaks. Note that any alternative results in a total energy efficiency loss for the OFDM waveform: the first by decreasing $\eta^{\rm d}$ and $\eta^{{\rm u} \textsc{f}}$, which increases $P_{\textsc{tx}}$ in \eqref{eq:TotalEE}, and the second by decreasing the achievable rates $\mathcal{R}^{\textsc{ul}}$ and $\mathcal{R}^{\textsc{dl}}$ due to the increased distortion levels imposed by clipping. This is one of the strongest motivations for using SC waveform. Further details about how the distortion levels imposed by clipping affects OFDM performance can be found in \cite{Tavares16}.

The CFO impact in maMIMO system performance has been discussed in \cite{Larsson15} and \cite{Mohammed16}. As presented in \cite{Moeneclaey95}, the CFO is another example of adverse effect much more severe for multicarrier systems, due to the longer duration of the OFDM symbol and the intercarrier interference due to the orthogonality loss. The impact of CFO in the total energy efficiency works as a performance-complexity trade-off: if a simple estimator is employed, the distortion imposed by carrier offset and/or phase noise decreases the achievable rates, while if a better estimator is used, the gains in achievable rates come at the price of either increased complexity or increased length of pilots. Note that both alternatives decrease the total energy efficiency. However, this trade-off is much more noticeable for OFDM than for SC, which constitutes another motivation for employing SC. Besides, an efficient method for improving CFO compensation is proposed in \cite{Mohammed16} for the TRMRC receiver operating in the maMIMO scenario, while also showing that the performance degradation due to imperfect CFO estimation diminishes with the increasing frequency selectivity of the channel, measured by the number of channel taps $L$.

The CP size also has a primary influence on the total energy efficiency of OFDM systems, but not in SC systems employing TRMRC which does not require CP. The minimum CP length is $N_{CP} = L-1$. While more extended CPs decrease the achievable rates of OFDM. Besides, for a given required CP length, one way of mitigating the impact of CP transmission on the achievable rates is increasing the number of subcarriers $N$. However, this strategy has an adverse effect on the PAPR, as discussed above. For the SC case, no CP is needed, as discussed in Section \ref{sec:SCwavef}, in such a way that only the number of channel taps $L$ impacts on the SC maMIMO system. While $L$ does not affect the achievable rates of TRMRC/SC, as can be seen in \eqref{eq:sinr_trmrc} and \eqref{eq:se_trmrc}, the computational complexity of such scheme depends on this parameter (see Table \ref{tab:complexity}), and therefore the total energy efficiency also decreases with $L$.

\section{Numerical Results}\label{sec:results}
\begin{figure}[!htbp]
\centering
\includegraphics[width=.859\textwidth]{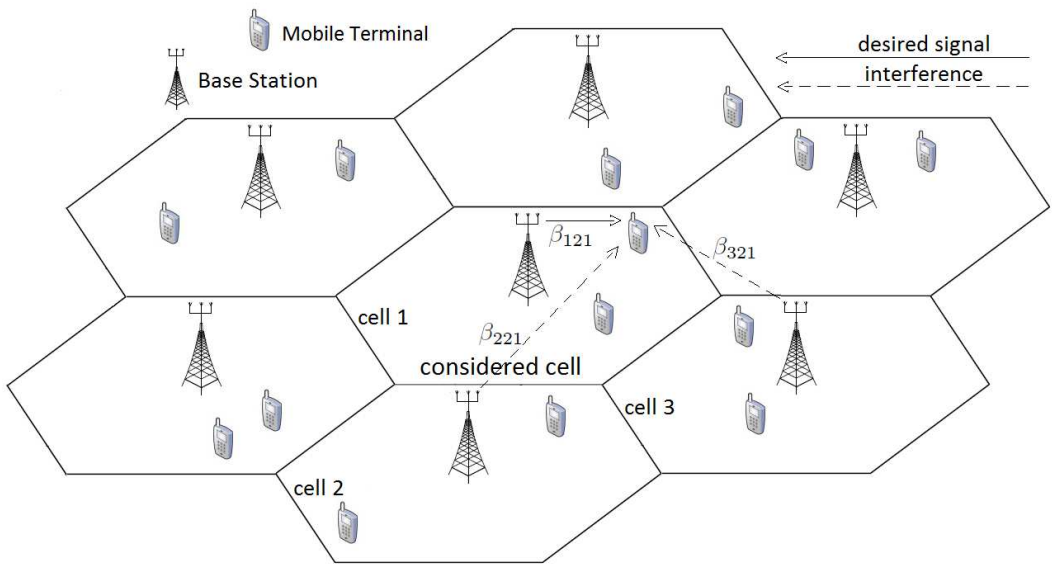}
\caption{Multicell maMIMO system model adopted.}
\label{fig:MC_maMIMO}
\end{figure}

Aiming at comparing the investigated schemes, we present in this Section numerical results for the maMIMO system performance under frequency-selective channels. We have adopted the same multicell scenario and parameters of \cite{Marinello17} with unitary frequency reuse factor, composed of $C = 7$ hexagonal cells of radius $r_{\rm cell} = 500$m, where $K$ users are uniformly distributed, except for a circle of 50m radius around the BS. The adopted multicell parameters are similar to that adopted in \cite{Marzetta16}, \cite{Debbah15}.  Each user is associated only with the BS which has the strongest long-term fading coefficient with him, while a uniform power allocation policy is assumed, both for simplicity. Besides, only the performance of the users at the central cell were computed, since they experience a more realistic interference. Unless otherwise specified, we have assumed $M = 100$ BS antennas, $N = 256$ subcarriers when OFDM is employed, a time-dispersive channel with $L = 16$, $\xi^{\rm d} = 0.6$, $\xi^{\rm u} = 0.4$, and an SNR of $0$dB for both pilots and UL data transmission. Moreover, to analyse the energy and spectral efficiencies behavior, we have varied the number of users $K$ in the Figures presented in this Section. The path loss decay exponent was assumed $\lambda = 3.8$, and the log-normal shadowing standard deviation as 8dB. The presented results were averaged over 100,000 spatial realizations for the users. Figure \ref{fig:MC_maMIMO} illustrates our adopted multicellular multiuser maMIMO system model.

\begin{figure}[!htbp]
\centering
\includegraphics[width=.7\textwidth]{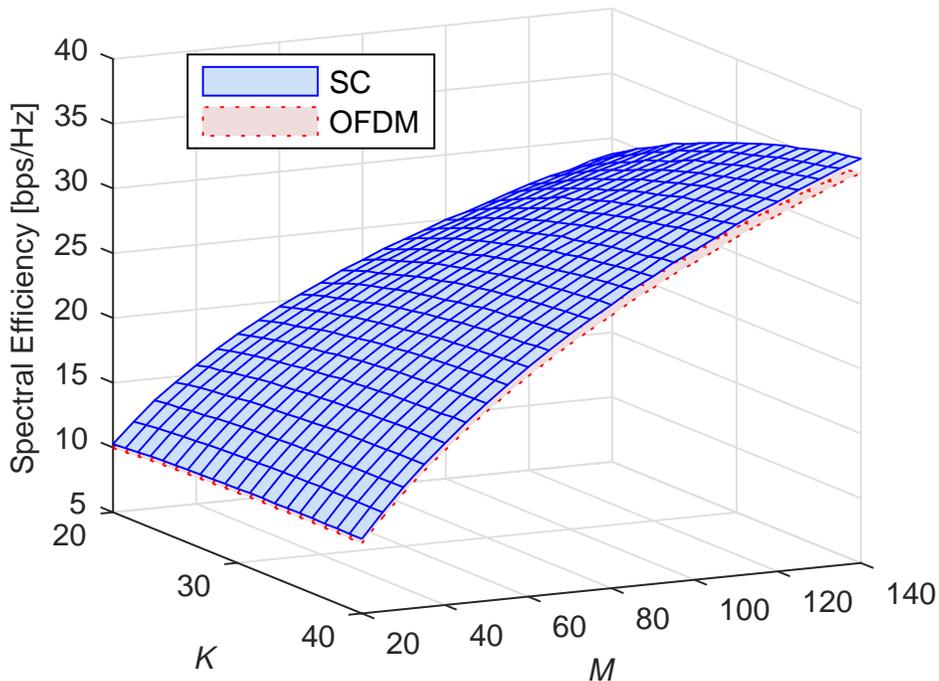}
\caption{Spectral efficiency {\emph{vs} $M$ and $K$ for} both investigated schemes.}
\label{fig:SE_compb}
\end{figure}

In \cite{Marinello17} it has been shown that the maMIMO system is able to operate with low UL SNR's, due to the large number of BS antennas, in such a way that the system performance saturates near 0dB. Moreover, both schemes have very similar SINR performance, but the achievable rates of TRMRC are slightly higher than that of FDMRC due to the inefficiency of the CP transmission, required by the latter in UL. Fig. \ref{fig:SE_compb} compares the UL and DL spectral efficiencies achieved by the TRMRC/SC {\it vs} FDMRC/OFDM schemes with increasing $M$ and $K$. Indeed, one can see that both have very close performances. This happens since both employ OFDM in DL, thus achieving the same spectral efficiency, while in UL the spectral efficiency of TRMRC/SC is slightly higher by not transmitting the CP. It has also been shown in [7] that the approximate expressions for SINR and the lower bounds for achievable rates are tight. While the error with respect to Monte-Carlo simulations of SINR was about 0.06 dB, the error between the simulated rates and the lower bound was about 0.075 bpcu.

\subsection{Total Energy Efficiency Evaluation}\label{sec:TotalEEcomp}

\begin{figure}[!htbp]
\centering
\includegraphics[width=.75\textwidth]{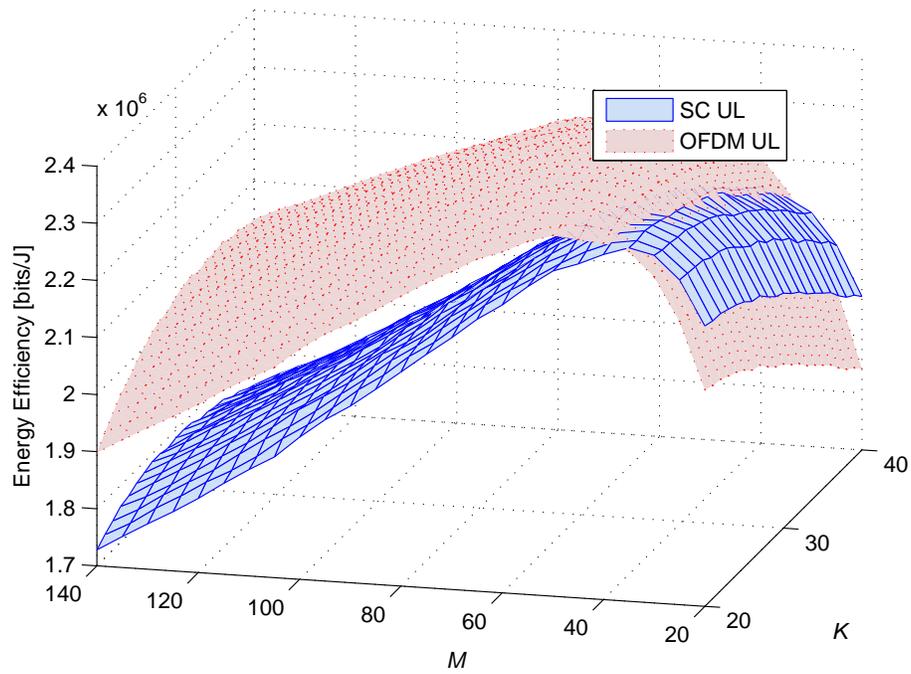}
\vspace{-5mm}
\caption{Total energy efficiency {\emph{vs} $M$ and $K$ for} both investigated schemes.}
\label{fig:TotalEE_compb}
\end{figure}

\begin{figure}[!htbp]
\centering
\includegraphics[width=.75\textwidth]{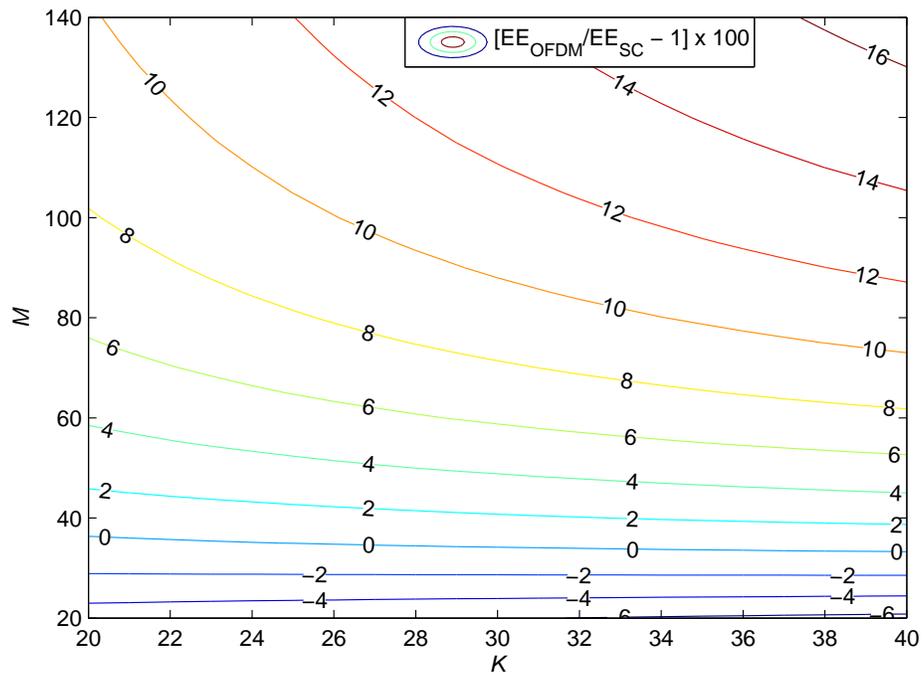}
\vspace{-4mm}
\caption{Contour plot showing the total energy efficiency gain (\%) of OFDM/FDMRC in comparison with SC/TRMRC.}
\label{fig:TotalEEGain}
\end{figure}

We evaluate in this subsection the total energy efficiencies of maMIMO systems employing either OFDM/FDMRC or SC/TRMRC as waveform/detector in the UL. The main parameters adopted in our simulations summarized in Table \ref{tab:parameters} are similar to those adopted in \cite{Debbah15}, \cite{Marzetta13}. {Fig.} \ref{fig:TotalEE_compb} depicts the total energy efficiencies of the investigated systems, with the variation of both number of BS antennas $M$ and number of users $K$. As can be seen, the maMIMO system employing FDMRC detector under OFDM waveform {has a higher} total energy efficiency than TRMRC under SC. The only exception occurs with low number of BS antennas ($M<35$), in which TRMRC/SC outperforms FDMRC/OFDM. However, in the massive MIMO scenario (typically with $M > 80$), the total energy efficiency of FDMRC/OFDM is higher. {Fig.} \ref{fig:TotalEEGain} corroborates this result, showing the total energy efficiency percentage gain of FDMRC/OFDM in comparison with TRMRC/SC. {As one can note from the figure, the total energy efficiency gain can achieve almost 20\% in the range of our investigation, and can be even higher for increased values of $M$ and $K$.}

\begin{figure}[!htbp]
\centering
\includegraphics[width=.75\textwidth]{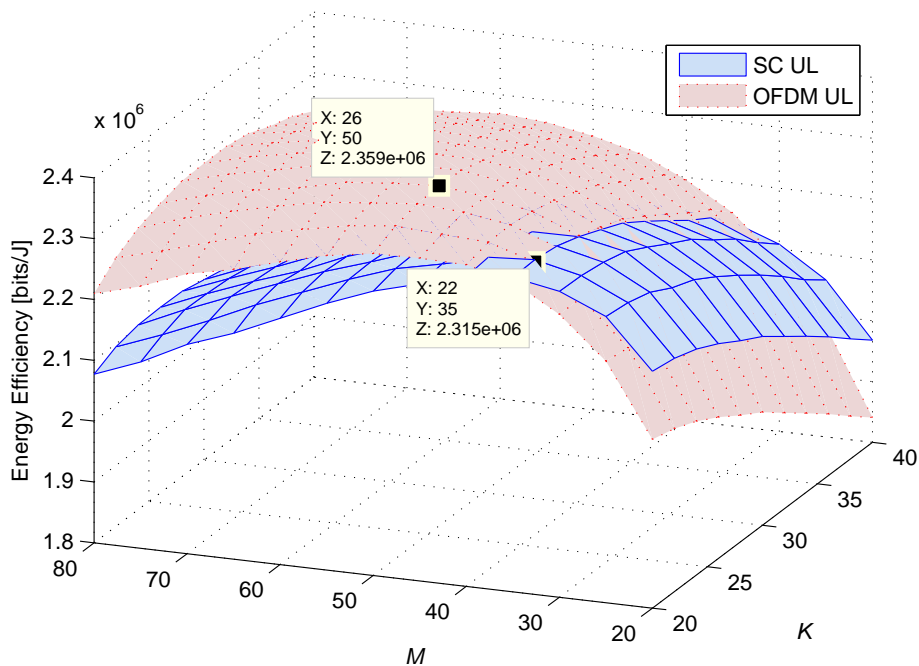}
\vspace{-5mm}
\caption{Total energy efficiency as a function of $M$ and $K$ (zoom of Fig. \ref{fig:TotalEE_compb}).}
\label{fig:TotalEE_zoom}
\end{figure}

Then, aiming to compare the best total energy efficiencies achieved by each scheme, we restrict the range of $K$ and $M$ in a shorter interval centered on their maxima. {Fig.} \ref{fig:TotalEE_zoom} {shows that} the optimal total energy efficiency achieved by TRMRC/SC is of 2.315 Mbits/J, which is attained with $M = 35$ antennas serving $K = 22$ users. On the other hand, the optimal operation point for FDMRC/OFDM is attained with $M = 50$ antennas serving $K = 22$ users, in which a total energy efficiency of 2.359 Mbits/J can be achieved. Therefore, FDMRC/OFDM is able to provide a total energy efficiency improvement of 0.044 Mbits/J with respect to TRMRC/SC, considering both schemes in their respective optimal total energy efficiency operation points.

\begin{figure}[!htbp]
\centering
\includegraphics[width=.75\textwidth]{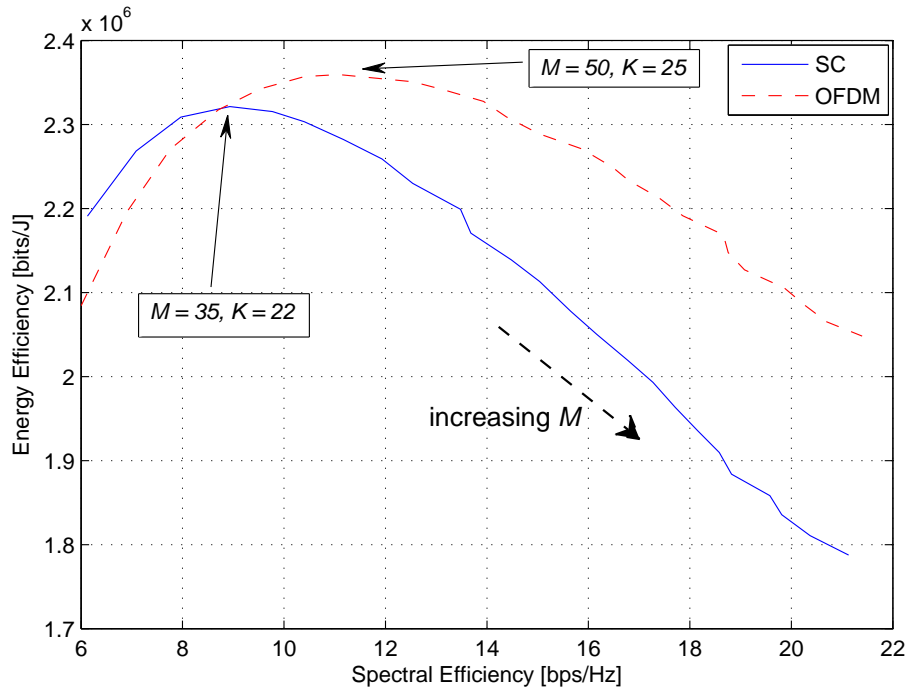}\\
\vspace{-2mm}
\caption{Total energy efficiency and sum spectral efficiency trade-off with increasing $M$.}
\label{fig:TotalEE_SE_tradeoff}
\end{figure}

Finally, Fig. \ref{fig:TotalEE_SE_tradeoff} shows the total energy efficiency and spectral efficiency trade-off for both investigated schemes. Each curve {shows} the total energy efficiency against spectral efficiency achieved for each scheme with the increasing number of BS antennas. As one can see from Fig. \ref{fig:TotalEE_compb}, for each number of BS antennas, there is a number of users $K$ which maximizes the total energy efficiency. For these operation points, we have collected their values of number of BS antennas, number of users, spectral and total energy efficiencies achieved. Then, it is depicted in Fig. \ref{fig:TotalEE_SE_tradeoff} the spectral efficiency and the total energy efficiency of such operation points. 
As one can infer, TRMRC/SC is a good alternative only for low number of BS antennas ($M<35$). Indeed, for higher values of $M$, FDMRC/OFDM {provides a higher} total energy efficiency under the same spectral efficiency, or alternatively {a higher} spectral efficiency under a target total energy efficiency. Notice that the spectral efficiency variation for each scheme in Fig. \ref{fig:TotalEE_SE_tradeoff} is mainly due to the different values of $M$ and $K$ along each curve.

\subsection{{Analysis with Decreasing Cell Radius and CP Length}}\label{sec:dec_Cell}

We analyse in this subsection the total energy efficiency of the investigated schemes with a decreasing cell radius. As long as the cell size diminishes, the CIR length $L$ also decreases, in such a way that a smaller CP can be employed. We start from $L = 16$, $r_{\rm cell} = 500$m, and $d_{\min} = 50$m, and gradually decrease $r_{\rm cell}$ and $d_{\min}$ in linear steps of 100 and 10, respectively, while $L$ is decreased by half in each step. Table \ref{tab:Perf_varL} shows the obtained results for the maximum total energy efficiency ${\rm EE_{max}}$, showing also the number of antennas, number of users and sum spectral efficiency (${\rm SE} = \sum_{k=1}^{K}\left( \mathcal{R}^{\textsc{ul}\star}_{k} + \mathcal{R}^{\textsc{dl}}_{k} \right)$) for which the maximum energy efficiency is obtained in each scenario. As one can see, TRMRC/SC achieves a higher total energy efficiency than FDMRC/OFDM when the cell radius decreases under 300m. This happens due to the lower value of $L$, which decreases the computational burden of TRMRC processing in Eq. \eqref{eq:est_data_trmrc}, while not changing the computational complexity of FDMRC. The table also shows that the maximum total energy efficiency operation point always occurs under a higher sum spectral efficiency for FDMRC/OFDM than for TRMRC/SC, due to the higher number of employed BS antennas and served users.

\begin{table}[!htbp]
\caption{Maximum Total Energy Efficiency with Decreasing Cell Radius for Each Scheme.}
\vspace{2mm}
\centering
\footnotesize
\begin{tabular}{|c|c|c|c|c|c|c|c|}
\hline
 \bf Scheme	 & \bf $r_{\rm cell}$ & \bf $d_{\min}$  & \bf $L$ & \bf ${\rm EE_{max}}$ & \bf {\rm SE} & \bf $M$ & \bf $K$ \\
           	 & [m]            &  [m]         &         &  [Mbits/J]        & [bps/Hz] &         &         \\
\hline
\hline
\multirow{5}{*}{\large{SC}} & 500 & 50 & 16 & 2.319 & 9.005 & 35 & 22 \\\cline{2-8}
 & 400 & 40 & 8 & 2.803 & 10.925 & 37 & 22 \\\cline{2-8}
 & 300 & 30 & 4 & \bf 3.351 & 13.753 & 40 & 25 \\\cline{2-8}
 & 200 & 20 & 2 & \bf 3.884 & 15.606 & 40 & 25 \\\cline{2-8}
 & 100 & 10 & 1 & \bf 4.218 & 17.060 & 41 & 27 \\\cline{2-8}
\hline
\hline
\multirow{5}{*}{\large{OFDM}} & 500 & 50 & 16 & \bf 2.361 & \bf 11.085 & 50 & 25 \\\cline{2-8}
 & 400 & 40 & 8 & \bf 2.831 & \bf 13.396 & 50 & 26 \\\cline{2-8}
 & 300 & 30 & 4 & 3.332 & \bf 15.473 & 49 & 25 \\\cline{2-8}
 & 200 & 20 & 2 & 3.786 & \bf 17.722 & 49 & 26 \\\cline{2-8}
 & 100 & 10 & 1 & 4.016 & \bf 18.628 & 47 & 28 \\\cline{2-8}
\hline
\end{tabular}
\label{tab:Perf_varL}
\end{table}
\normalsize

\subsection{{Analysis with Increasing Computational Efficiency}}\label{sec:inc_CE}

\begin{figure}[!htbp]
\centering
\includegraphics[width=.75\textwidth]{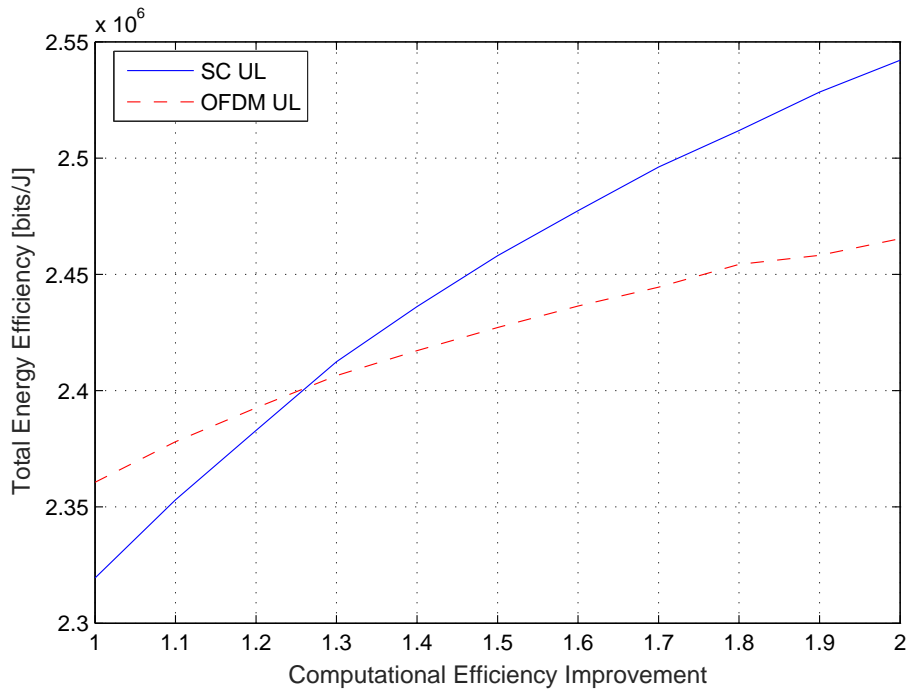}
\vspace{-5mm}
\caption{Total energy efficiency with increasing computational efficiency (both $\mathcal{L}_{\textsc{bs}}$ and $\mathcal{L}_{\textsc{mt}}$), for $L$=16 and $r_{\rm{ext}} = 500$.}
\label{fig:TotalEE_CEI}
\end{figure}

Our previous results demonstrated that TRMRC/SC achieves a higher sum rate than FDMRC/OFDM, while spending less power with the irradiated signal and power amplifier. However, the higher computational complexity of TRMRC processing makes its total energy efficiency lower than that of FDMRC/OFDM. In this subsection we conjecture that such situation may change in few years, since the evolution of computers and processors tends to improve the computational efficiency parameters, $\mathcal{L}_{\textsc{bs}}$ and $\mathcal{L}_{\textsc{mt}}$. Therefore, the power required to run the TRMRC processing would decrease, increasing its total energy efficiency. Fig. \ref{fig:TotalEE_CEI} shows the maximum total energy efficiency achieved by both schemes with increasing computational efficiencies, \emph{i.e.}, the $\mathcal{L}_{\textsc{bs}}$ and $\mathcal{L}_{\textsc{mt}}$ in Table \ref{tab:parameters} were scaled by the factor represented in the horizontal axis. As depicted in the figure, a 30\% improvement in computational efficiencies is sufficient to make the total energy efficiency of TRMRC/SC higher than that of FDMRC/OFDM. Therefore, employing TRMRC under SC waveform in the UL of maMIMO systems may become a promising alternative in a near future aiming to implement very energy efficient communication systems.

\section{Conclusion}\label{sec:concl}

In this work, we have evaluated the total energy efficiency of maMIMO systems under frequency-selective fading channels in the presence of pilot contamination. We have compared the conventional FDMRC receiver operating with OFDM versus the TRMRC receiver operating under SC waveform. Our choice for the total energy efficiency metric is due to its comprehensiveness, and ability of incorporating several different aspects in a unified analysis. {Our results demonstrated the superiority of FDMRC receiver under OFDM waveform in terms of total energy efficiency. Despite of the higher sum rates achieved by TRMRC/SC, and the lower power expenditures related to the transmitted signal and power amplifier, the higher computational complexity required to run the TRMRC processing makes the total energy efficiency of this scheme lower than that achieved by FDMRC under OFDM. However, this complexity is highly dependent on the CIR length $L$, in such a way that if a shorter cell size is considered, TRMRC/SC becomes a better alternative in terms of total energy efficiency. Besides, the power required to run the detection algorithms decreases if the computational efficiency of the available equipments increase. Our analysis have indicated that a 30\% increase on the computational efficiency makes the TRMRC/SC scheme more efficient than FDMRC/OFDM. Therefore, the latter topology could be a promising alternative for implementation in the UL of maMIMO systems in the next few years.}

\section*{Acknowledgements}
This work was supported in part by the National Council for Scientific and Technological Development (CNPq) of Brazil under Grants 304066/2015-0 and 404079/2016-4; in part by Londrina State University, Paraná State Government (UEL); and in part by Escola Politécnica of the University of São Paulo (EPUSP).


\begin{thebibliography}{10}
\providecommand{\url}[1]{#1}
\csname url@samestyle\endcsname
\providecommand{\newblock}{\relax}
\providecommand{\bibinfo}[2]{#2}
\providecommand{\BIBentrySTDinterwordspacing}{\spaceskip=0pt\relax}
\providecommand{\BIBentryALTinterwordstretchfactor}{4}
\providecommand{\BIBentryALTinterwordspacing}{\spaceskip=\fontdimen2\font plus
\BIBentryALTinterwordstretchfactor\fontdimen3\font minus
  \fontdimen4\font\relax}
\providecommand{\BIBforeignlanguage}[2]{{%
\expandafter\ifx\csname l@#1\endcsname\relax
\typeout{** WARNING: IEEEtran.bst: No hyphenation pattern has been}%
\typeout{** loaded for the language `#1'. Using the pattern for}%
\typeout{** the default language instead.}%
\else
\language=\csname l@#1\endcsname
\fi
#2}}
\providecommand{\BIBdecl}{\relax}
\BIBdecl

\bibitem{Marzetta16}
T.~Marzetta, E.~Larsson, H.~Yang, and H.~Ngo, \emph{Fundamentals of Massive
  MIMO}.\hskip 1em plus 0.5em minus 0.4em\relax New York, NY, USA: Cambridge
  University Press, 2016.

\bibitem{Ngo13_EE_SE}
H.~Q. Ngo, E.~Larsson, and T.~Marzetta, ``Energy and spectral efficiency of
  very large multiuser {MIMO} systems,'' \emph{IEEE Transactions on
  Communications}, vol.~61, no.~4, pp. 1436--1449, April 2013.

\bibitem{Marzetta10}
T.~Marzetta, ``Noncooperative cellular wireless with unlimited numbers of base
  station antennas,'' \emph{IEEE Transactions on Wireless Communications},
  vol.~9, no.~11, pp. 3590--3600, 2010.

\bibitem{Debbah16}
E.~Björnson, E.~G. Larsson, and M.~Debbah, ``Massive {MIMO} for maximal
  spectral efficiency: How many users and pilots should be allocated?''
  \emph{IEEE Transactions on Wireless Communications}, vol.~15, no.~2, pp.
  1293--1308, Feb 2016.

\bibitem{Zhang15}
Q.~Zhang, S.~Jin, M.~McKay, D.~Morales-Jimenez, and H.~Zhu, ``{Power Allocation
  Schemes for Multicell Massive MIMO Systems},'' \emph{IEEE Transactions on
  Wireless Communications}, vol.~14, no.~11, pp. 5941--5955, Nov 2015.

\bibitem{Falconer02}
D.~Falconer, S.~L. Ariyavisitakul, A.~Benyamin-Seeyar, and B.~Eidson,
  ``Frequency domain equalization for single-carrier broadband wireless
  systems,'' \emph{IEEE Communications Magazine}, vol.~40, no.~4, pp. 58--66,
  Apr 2002.

\bibitem{Marinello17}
J.~C.~M. Filho, C.~Panazio, and T.~Abrão, ``Uplink performance of
  single-carrier receiver in massive {MIMO} with pilot contamination,''
  \emph{IEEE Access}, vol.~5, pp. 8669--8681, 2017.

\bibitem{GYLi_Survey_14}
L.~Lu, G.~Y. Li, A.~L. Swindlehurst, A.~Ashikhmin, and R.~Zhang, ``An overview
  of massive {MIMO}: Benefits and challenges,'' \emph{IEEE Journal of Selected
  Topics in Signal Processing}, vol.~8, no.~5, pp. 742--758, Oct 2014.

\bibitem{Benvenuto02}
N.~Benvenuto and S.~Tomasin, ``On the comparison between {OFDM} and single
  carrier modulation with a {DFE} using a frequency-domain feedforward
  filter,'' \emph{IEEE Transactions on Communications}, vol.~50, no.~6, pp.
  947--955, Jun 2002.

\bibitem{Tomasin05}
S.~Tomasin, ``Overlap and save frequency domain {DFE} for throughput efficient
  single carrier transmission,'' in \emph{2005 IEEE 16th International
  Symposium on Personal, Indoor and Mobile Radio Communications}, vol.~2, Sept
  2005, pp. 1199--1203 Vol. 2.

\bibitem{Larsson15}
A.~Pitarokoilis, S.~K. Mohammed, and E.~G. Larsson, ``Uplink performance of
  time-reversal {MRC} in massive {MIMO} systems subject to phase noise,''
  \emph{IEEE Transactions on Wireless Communications}, vol.~14, no.~2, pp.
  711--723, Feb 2015.

\bibitem{Larsson12}
------, ``On the optimality of single-carrier transmission in large-scale
  antenna systems,'' \emph{IEEE Wireless Communications Letters}, vol.~1,
  no.~4, pp. 276--279, August 2012.

\bibitem{Mohammed16}
S.~Mukherjee and S.~K. Mohammed, ``{Impact of Frequency Selectivity on the
  Information Rate Performance of CFO Impaired Single-Carrier Massive MU-MIMO
  Uplink},'' \emph{IEEE Wireless Communications Letters}, vol.~5, no.~6, pp.
  648--651, Dec 2016.

\bibitem{Dinis18}
Z.~Mokhtari, M.~Sabbaghian, and R.~Dinis, ``{Massive MIMO Downlink Based on
  Single Carrier Frequency Domain Processing},'' \emph{IEEE Transactions on
  Communications}, vol.~66, no.~3, pp. 1164--1175, March 2018.

\bibitem{Song18}
Y.~Sun, J.~Wang, L.~He, and J.~Song, ``{Uplink Spectral Efficiency Analysis and
  Optimization for Massive SC-SM MIMO With Frequency Domain Detection},''
  \emph{IEEE Transactions on Vehicular Technology}, vol.~67, no.~5, pp.
  3937--3949, May 2018.

\bibitem{He18}
Y.~Sun, J.~Wang, and L.~He, ``{Spectral Efficiency Enhancement With Power
  Allocation for Massive SC-SM MIMO Uplink},'' \emph{IEEE Communications
  Letters}, vol.~22, no.~1, pp. 101--104, Jan 2018.

\bibitem{Heath17}
C.~Mollén, J.~Choi, E.~G. Larsson, and R.~W. Heath, ``{Uplink Performance of
  Wideband Massive MIMO With One-Bit ADCs},'' \emph{IEEE Transactions on
  Wireless Communications}, vol.~16, no.~1, pp. 87--100, Jan 2017.

\bibitem{Debbah15}
E.~Björnson, L.~Sanguinetti, J.~Hoydis, and M.~Debbah, ``Optimal design of
  energy-efficient multi-user {MIMO} systems: Is massive {MIMO} the answer?''
  \emph{IEEE Transactions on Wireless Communications}, vol.~14, no.~6, pp.
  3059--3075, June 2015.

\bibitem{Marzetta13}
H.~Yang and T.~L. Marzetta, ``{Total energy efficiency of cellular large scale
  antenna system multiple access mobile networks},'' in \emph{2013 IEEE Online
  Conference on Green Communications (OnlineGreenComm)}, Oct 2013, pp. 27--32.

\bibitem{Yang17}
K.~Yang, S.~Martin, D.~Quadri, J.~Wu, and G.~Feng, ``{Energy-Efficient Downlink
  Resource Allocation in Heterogeneous OFDMA Networks},'' \emph{IEEE
  Transactions on Vehicular Technology}, vol.~66, no.~6, pp. 5086--5098, June
  2017.

\bibitem{Fu17}
S.~Fu, H.~Wen, J.~Wu, and B.~Wu, ``{Energy-Efficient Precoded Coordinated
  Multi-Point Transmission With Pricing Power Game Mechanism},'' \emph{IEEE
  Systems Journal}, vol.~11, no.~2, pp. 578--587, June 2017.

\bibitem{Liao17}
J.~An, K.~Yang, J.~Wu, N.~Ye, S.~Guo, and Z.~Liao, ``{Achieving Sustainable
  Ultra-Dense Heterogeneous Networks for 5G},'' \emph{IEEE Communications
  Magazine}, vol.~55, no.~12, pp. 84--90, December 2017.

\bibitem{Xin16}
Y.~Xin, D.~Wang, J.~Li, H.~Zhu, J.~Wang, and X.~You, ``{Area Spectral
  Efficiency and Area Energy Efficiency of Massive {MIMO} Cellular Systems},''
  \emph{IEEE Transactions on Vehicular Technology}, vol.~65, no.~5, pp.
  3243--3254, May 2016.

\bibitem{Golub96}
G.~H. Golub and C.~F.~V. Loan, \emph{Matrix Computations}.\hskip 1em plus 0.5em
  minus 0.4em\relax Maryland, USA: Johns Hopkins University Press, 1996.

\bibitem{Horowitz17}
A.~Pedram, S.~Richardson, M.~Horowitz, S.~Galal, and S.~Kvatinsky, ``{Dark
  Memory and Accelerator-Rich System Optimization in the Dark Silicon Era},''
  \emph{IEEE Design Test}, vol.~34, no.~2, pp. 39--50, April 2017.

\bibitem{Horowitz14}
M.~Horowitz, ``1.1 computing's energy problem (and what we can do about it),''
  in \emph{{2014 IEEE International Solid-State Circuits Conference Digest of
  Technical Papers (ISSCC)}}, Feb 2014, pp. 10--14.

\bibitem{Bahai04}
S.~Cui, A.~J. Goldsmith, and A.~Bahai, ``Energy-efficiency of {MIMO} and
  cooperative {MIMO} techniques in sensor networks,'' \emph{IEEE Journal on
  Selected Areas in Communications}, vol.~22, no.~6, pp. 1089--1098, Aug 2004.

\bibitem{Tavares16}
C.~H.~A. Tavares, J.~C.~M. Filho, C.~M. Panazio, and T.~Abrão, ``{Input
  Back-Off Optimization in OFDM Systems Under Ideal Pre-Distorters},''
  \emph{IEEE Wireless Communications Letters}, vol.~5, no.~5, pp. 464--467, Oct
  2016.

\bibitem{Moeneclaey95}
T.~Pollet, M.~V. Bladel, and M.~Moeneclaey, ``{BER sensitivity of OFDM systems
  to carrier frequency offset and Wiener phase noise},'' \emph{IEEE
  Transactions on Communications}, vol.~43, no. 2/3/4, pp. 191--193, Feb 1995.

\end{thebibliography}
\end{document}